\documentclass[sigconf,nonacm]{acmart}

\AtBeginDocument{%
  }
    
\usepackage{color, colortbl}
\usepackage{tkz-euclide}
\usepackage{pgfplots}
\usepackage{url}
\usepackage{pifont}
\usepackage{pgf-pie}
\usepackage{framed}
\definecolor{yesgreen}{HTML}{4DAF4A}
\definecolor{nored}{HTML}{E41A1C}
\definecolor{dkyellow}{HTML}{FFD92F}

\usepackage{xcolor}
\usepackage{array}

\definecolor{yesgreen}{HTML}{4DAF4A}
\definecolor{nored}{HTML}{E41A1C}
\definecolor{dkyellow}{HTML}{FFD92F}

\newcommand{\cmark}{\ding{51}} 
\newcommand{\xmark}{\ding{55}} 

\pgfplotsset{compat=1.18}

\newif\ifstatus
\statustrue 

\definecolor{DarkGreen}{RGB}{0,120,0}
\renewcommand{\cmark}{{\color{DarkGreen}\ding{51}}}
\renewcommand{\xmark}{{\color{red}\ding{55}}}

\setcopyright{acmlicensed}
\copyrightyear{2025}
\acmYear{2025}

\acmConference[Conference acronym]{Conference Name}{October, 2025}{Location}
\acmISBN{978-1-4503-XXXX-X/18/06}

\begin{document}

\title{Benchmarking Large Language Models for IoC Recovery under Adversarial Code Obfuscation and Encryption}
\author{Jaime Morales}
\affiliation{%
  \institution{Universidad Carlos III}
  \city{Madrid}
  \country{Spain}
}
\email{100277258@alumnos.uc3m.es}

\author{Sergio Pastrana}
\affiliation{%
  \institution{Universidad Carlos III}
  \city{Madrid}
  \country{Spain}
}
\email{spastran@inf.uc3m.es}

\author{Juan Tapiador}
\affiliation{%
  \institution{Universidad Carlos III}
  \city{Madrid}
  \country{Spain}
}
\email{jestevez@inf.uc3m.es}

\renewcommand{\shortauthors}{Morales et al.}

\begin{abstract}
Software obfuscation and encryption present persistent challenges for program comprehension and security analysis, particularly when adversaries conceal Indicators of Compromise (IoCs) such as IP addresses within source code. While Large Language Models (LLMs) have recently demonstrated remarkable progress in code reasoning and transformation, their resilience against adversarial concealment techniques remains largely uncharted.

This paper introduces a systematic benchmark for secret detection under adversarial code transformations, designed to evaluate the capacity of LLMs to recover IoCs embedded in obfuscated and encrypted JavaScript programs. We construct a dataset of 336 programs, progressively transformed through 12 levels of obfuscation and cryptographic concealment (including XOR and AES-256), to emulate realistic threat scenarios. An automated evaluation framework standardizes LLM queries and responses, enabling reproducible, large-scale testing across diverse models.

Our results reveal a dichotomy: while LLMs exhibit high success against lightweight transformations such as variable renaming and Base64 encoding, encryption-based concealment severely degrades detection performance. These findings establish encryption as a critical frontier for LLM-driven code analysis and highlight both current limitations and avenues for advancing automated threat intelligence.
\end{abstract}



\maketitle

\section{Introduction}

Code obfuscation aims to deliberately conceal the analysis of source code, either by humans or machines. It remains a fundamental challenge for cybersecurity analysts since it is widely used by adversaries to conceal malicious intent and hinder analysis. Techniques such as renaming identifiers, control-flow manipulation, or string encoding, complicate both human inspection and automated analysis pipelines. One such analyses consist of extracting threat intelligence from malicious sources, commonly in form of Indicators of Compromise (IoCs) such as IP addresses, contacted domains, or embedded API keys. These IoCs are also obfuscated to evade detection. This directly impacts Cyber Threat Intelligence (CTI) workflows, where timely and accurate extraction of IoCs is critical for threat hunting, intrusion detection, and incident response. The inability of automated systems to reliably extract IoCs from obfuscated code weakens defensive capabilities and slows down response times.

Recent advances in Large Language Models (LLMs) have demonstrated strong capabilities for understanding code semantics, including deobfuscation and vulnerability, making them an emergent tool for security analysis \cite{liang2025breakingobfuscationclusterawaregraph} \cite{jiang2025cascadellmpoweredjavascriptdeobfuscator}. However, current state-of-the-art approaches still fall short when applied to the systematic detection of IoCs under progressive obfuscation. Benchmarking efforts have primarily measured structural recovery or readability improvements, rather than focusing on adversarial concealment of security-relevant artifacts. Moreover, existing works that evaluate JavaScript deobfuscation with LLMs do not consider encryption techniques~\cite{chen2025jsdeobsbench}, and also do not focus on the automatic extraction of IoCs. This leaves two critical research questions:
\begin{itemize}
    \item \textbf{RQ1: Are LLMs capable of finding security-relevant artifacts in obfuscated source code?}
    \item \textbf{RQ2: Do different obfuscation techniques affect the ability of LLMs to accurately extract security-relevant artifacts from obfuscated source code?}
\end{itemize} 

In this paper, we address these questions by proposing a methodology to analyze the effectiveness of LLMs in identifying IoCs in obfuscated code at scale. Our research is guided by three overarching goals.

\begin{enumerate}
    \item Realistic adversarial scenarios. We consider an experimental environment with realistic concealment techniques, by applying 12 distinct obfuscation and encryption transformations, ranging from lightweight syntactic changes to robust encryption with AES-256. The goal is to simulate diverse adversarial conditions and evaluate the capacity of LLMs to handle them.
    
    \item Reproducible evaluation. We build a modular and automated framework capable of systematically querying multiple LLMs, normalizing their responses into a unified schema, and enabling fair, large-scale cross-model comparison under controlled conditions.
    
    \item Error and capability characterization. We systematically analyze the performance of state-of-the-art LLMs, identifying their strengths in handling lightweight transformations and their consistent limitations when confronted with encrypted or semantically distorted code.
\end{enumerate}

Based on these goals, we developed a fully automated methodology that integrates multi-level obfuscation on an initial set of source code samples and performs scalable evaluation using several commercial LLMs. This pipeline enables realistic and repeatable testing of model capabilities under increasing complexity.

Building upon this design, the main contributions of this paper are as follows:

\begin{enumerate}
    \item We present an automated and reproducible methodology to evaluate multiple LLMs (ChatGPT, Gemini, Anthropic, Grok, and Cohere) on their ability to detect hidden Indicators of Compromise (IoCs) in source code subjected to different obfuscation levels.
    
    \item We construct and release a large-scale dataset derived from 336 original JavaScript samples, which were systematically transformed using 12 obfuscation techniques, resulting in 4,368 modified instances suitable for benchmarking.
    
    \item We provide a comprehensive empirical evaluation comparing multiple LLMs across obfuscation phases, highlighting detection accuracy, reliability, and hallucination patterns. These results offer new insights into how LLMs interpret obfuscated code and reveal their limitations in real-world security contexts.
\end{enumerate}

The rest of the paper is structured as follows. Section \ref{sec:sota} describes the background and reviews existing research on LLM-based code analysis and secret detection. Section \ref{sec:methodology}  defines the metrics used to evaluate LLMs and presents the dataset of modified JavaScript code samples. Section \ref{sec:experimentation} describes the experimental setup, testing methodology, and LLM query process. 
Finally, section \ref{sec:discussion} explores key findings, challenges, and limitations of using LLMs for IoC extraction, and Section \ref{sec:conclusion} concludes the paper. 

\section{Background}
\label{sec:sota}


The integration of Large Language Models (LLMs) into cybersecurity enables new capabilities for automating threat detection and intelligence collection. Concretely, a a key objective in Cyber Threat Intelligence (CTI) is the identification and extraction of Indicators of Compromise (IoCs)—such as IP addresses, domains, or file hashes—that reveal traces of adversarial activity. Automating this process across multiple data sources greatly enhances the scalability and timeliness of CTI operations, reducing the reliance on manual analysis~\cite{10.1145/2976749.2978315}. While prior research has primarily focused on extracting IoCs from unstructured text, such as security reports~\cite{caballero2023rise} or hacking forums~\cite{mischinger2024ioc}, additional challenges arise when these indicators are concealed within program code through obfuscation or encryption. In this context, recent studies have explored how LLMs can support CTI tasks for intrusion detection, malware analysis, and code deobfuscation, suggesting their potential to advance automated IoC extraction in adversarial environments \cite{alkaraki2024exploringllmsmalwaredetection,10.1145/2976749.2978315}.

We next describe the main types of obfuscation techniques:

\begin{itemize}
\item \textbf{Control Flow Obfuscation:} This technique hides the actual execution flow of a program, making static and dynamic analysis more complex. Common examples include adding invalid functions or ``dead code'', jumps, and redirection that complicate understanding the code.

\item Function Obfuscation: Functions in the code can be modified so that their purpose is not easily determined, especially in static code analysis. Some of the most common techniques include renaming functions and variables with random names that are non-descriptive.

\item Data Obfuscation. This category seeks to hide key data in the code or malware, such as IP addresses, strings, encryption keys, and any other sensitive data that an analyst could easily identify. This includes simple obfuscation techniques, such as Base64 encoding or XOR encryption.

\item Cryptographic Obfuscation. It relies on advanced cryptography algorithms, such as AES-256, which makes it much harder to retrieve the decrypted data without the symmetric key.


\item Polymorphic and Metamorphic Obfuscation. These are techniques that change the code each time it runs, to avoid signature-based or behavior-based detection.
\begin{itemize}
    \item Polymorphism: Each instance of the malware has different code, even if it performs the same actions. Code generators can be applied to create variations of the malware.
    \item Metamorphism: The malware rewrites itself upon each execution, meaning even if an instance is captured, it will not be useful for future analysis, as it will be completely different the next time it runs.
\end{itemize}
\end{itemize}

\begin{table}[h]
    \caption{Types of obfuscation used in the study}
    \label{table:obfuscation1}
    \centering
    \renewcommand{\arraystretch}{1.2} 
    \begin{tabular}{lc}
        \toprule
        Obfuscation Type & Used \\ \midrule
        Control Flow Obfuscation & Yes \\
        Data Obfuscation & Yes \\
        Cryptographic Obfuscation & Yes \\
        Function Obfuscation & Yes \\
        Polymorphic and Metamorphic Obfuscation & No \\ \bottomrule
    \end{tabular}
\end{table}

As shown in Table 1, this study includes all the obfuscation types except for polymorphism and metamorphism mechanisms. The goal is to evaluate whether current LLMs can reliably identify concealed IoCs within obfuscated code. Concretely, we assess their ability to recognize the presence of hidden artifacts, and when possible, to retrieve the actual value. The evaluation is performed under a controlled and reproducible framework, designed to systematically measure the detection performance and consistency of model responses.

\subsection{Related Work}

The application of Large Language Models (LLMs) in cybersecurity has progressed from initial feasibility studies to comprehensive benchmarking and operational deployment. Early work primarily explored capabilities such as code deobfuscation and threat intelligence extraction. Ramesh et al. systematically evaluated the autonomous extraction of Indicators of Compromise (IoCs) from obfuscated and encrypted samples, illustrating the potential of LLMs as cyber defenders able to perform sophisticated threat intelligence tasks in adversarial contexts~\cite{ramesh2025llmscyberdefenders}. Zhang et al. introduced ObfBench, a large-scale benchmark specifically designed to assess the ability of both machine learning models and LLMs to deobfuscate code across diverse obfuscation strategies, thereby advancing standardized evaluations within the field~\cite{chen2025jsdeobsbench}.

Regarding code deobfuscation, which is the focus of this paper, the work by Patsakis et al. analyzed LLM performance on obfuscated scripts from the Emotet malware family~\cite{patsakis2024assessingllmsmaliciouscode}. Their study demonstrated that, although LLMs could not resolve all obfuscation layers, they were able to uncover hidden strings and simplify structural elements, thereby assisting human analysts. Its key contribution lies in validating LLMs as \textit{auxiliary forensic tools} for malware investigation.

In a parallel work to ours, Alkari et al. provided a broad survey of LLM applications in malware detection~\cite{alkaraki2024exploringllmsmalwaredetection}. Beyond cataloging use cases, the authors proposed a structured framework for integrating LLMs into detection pipelines and, importantly, emphasized dual-use risks. Their work advanced the discussion from isolated case studies to \textit{system-level integration and governance considerations}.

In 2025 the focus shifted toward benchmarks and production-ready systems. \cite{chen2025jsdeobsbench} introduced a large-scale benchmark that systematically evaluates LLMs on syntactic simplification, readability, and execution correctness. The benchmark highlighted strengths of frontier models such as GPT-4o, but also revealed persistent limitations in semantic preservation. Its improvement over earlier studies lies in providing a \textit{reproducible, comparative evaluation framework} rather than ad hoc experiments. In parallel, Choi et al. combine fine-tuning and structured prompt engineering over C/C++ programs obfuscated with Tigress and OLLVM, and report consistent gains in deobfuscation quality and robustness across multiple obfuscation levels~\cite{choi2026towarddeobfuscationllm}. This reinforces the need for standardized, multi-level evaluation rather than one-off case studies.

Complementing this, \textit{CASCADE} \cite{jiang2025cascadellmpoweredjavascriptdeobfuscator} presented a hybrid approach combining Gemini with compiler-based intermediate representations (JSIR). CASCADE achieved near-perfect detection of obfuscation prelude functions and successfully recovered thousands of string literals with high accuracy and throughput. Crucially, the system has been deployed in production at Google, marking a transition from academic prototypes to \textit{industrial-grade, scalable deobfuscation solutions}. 
\textit{DeCoda} is a framework that integrates LLM prompting with cluster-aware graph learning for deobfuscation~\cite{liang2025breakingobfuscationclusterawaregraph}. By leveraging structural code properties alongside LLM reasoning, DeCoda achieved detection accuracy exceeding 94\%. Its novelty lies in demonstrating the benefits of \textit{hybrid symbolic-neural approaches}, balancing robustness and generalization.

\begin{framed}
\textbf{Analysis and research gap. }
Existing studies trace the evolution of LLM research in cybersecurity: from early explorations of feasibility, through structured benchmarks, to production-ready systems. They collectively demonstrate the potential of LLMs for code analysis, secret recovery, and malware detection, while also highlighting the need for systematic evaluation, reproducibility, and careful consideration of risks. Table~\ref{tab:related_matrix} summarizes the main characteristics of these works along several dimensions, and contrasts them with the scope of this paper.
Despite these advances, existing works primarily concentrate on structural code simplification, string recovery, or benchmark-scale evaluations of obfuscation techniques. A critical gap remains in the systematic study of \textit{secret detection under cryptographic concealment}, where adversaries embed Indicators of Compromise (IoCs) such as IP addresses within obfuscated or encrypted code. Addressing this gap requires not only progressively designed benchmarks but also evaluation frameworks capable of capturing the specific challenges posed by encryption-based transformations. In the following section, we present our methodology to construct such a benchmark and framework, focusing on the progressive concealment of IoCs across multiple obfuscation and encryption levels.
\end{framed}

\begin{table}[t]
\caption{Comparison between this paper and related work.}
\label{tab:related_matrix}
\centering
\renewcommand{\arraystretch}{1.15}
\resizebox{\columnwidth}{!}{%
\begin{tabular}{lccccccc}
\toprule
Feature &
\cite{patsakis2024assessingllmsmaliciouscode} &
\cite{alkaraki2024exploringllmsmalwaredetection} &
\cite{chen2025jsdeobsbench} &
\cite{jiang2025cascadellmpoweredjavascriptdeobfuscator} &
\cite{liang2025breakingobfuscationclusterawaregraph} &
\cite{choi2026towarddeobfuscationllm} &
This Paper \\
\midrule
Systematic benchmark & \xmark & \xmark & \cmark & \xmark & \xmark & \cmark & \cmark \\
JavaScript focus & \cmark & \xmark & \cmark & \cmark & \cmark & \xmark & \cmark \\
Hybrid / ML-based approach & \cmark & \cmark & \cmark & \cmark & \cmark & \cmark & \cmark \\
IoC detection focus & \xmark & \xmark & \xmark & \xmark & \xmark & \xmark& \textbf{\cmark} \\
Use of encryption & \xmark & \xmark & \xmark & \xmark & \xmark & \xmark & \textbf{\cmark} \\
Progressive obfuscation levels & \xmark & \xmark & \xmark & \xmark & \xmark & \cmark & \textbf{\cmark} \\
Automated evaluation pipeline & \xmark & \xmark & \cmark & \cmark & \xmark & \cmark & \textbf{\cmark} \\
Hallucination analysis & \xmark & \xmark & \xmark & \xmark & \xmark & \xmark & \textbf{\cmark} \\
CTI relevance (IoC recovery) & \xmark & \xmark & \xmark & \xmark & \xmark & \xmark & \textbf{\cmark} \\
\bottomrule
\end{tabular}
}
\end{table}

\section{Methodology}
\label{sec:methodology}

\begin{figure*}[t]
    \centering
    \includegraphics[width=\textwidth]{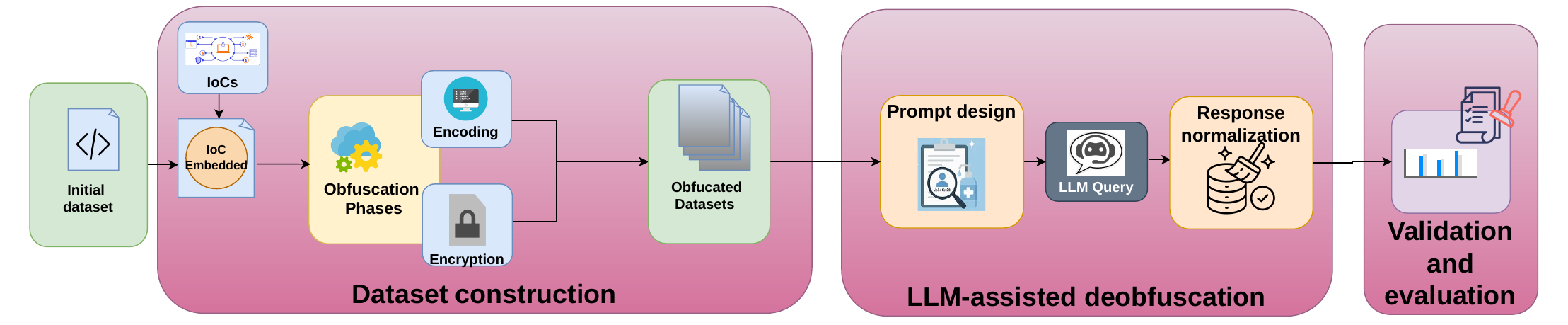} 
    \caption{Workflow of the proposed system integrating obfuscation, AI-assisted deobfuscation, and automated evaluation.}
    \label{fig:architecture}
\end{figure*}

Building on the gap identified in the literature we designed a reproducible and automated methodology to assess the ability of LLMs to locate and recover Indicators of Compromise (IoCs) embedded in obfuscated and encrypted code. Our methodology is depicted in Figure~\ref{fig:architecture}, and consists of two main components. 

First, the \textit{Dataset Construction} aims to build a benchmark that reflects realistic adversarial scenarios. Starting from a clean set of benign, executable programs, we apply a predefined sequence of transformations—ranging from simple encodings to symmetric encryption schemes—to produce 12 progressive concealment levels. Each level is designed to incrementally increase the difficulty of IoC recovery, while preserving traceability between the original program and its transformed variants, thus enabling objective ground-truth verification.

Second, the \textit{AI-assisted Deobfuscation} defines an automated pipeline to interrogate multiple models under consistent experimental conditions. This pipeline includes: (i) controlled and standardized prompt templates; (ii) a unified response schema to enable automated evaluation
; (iii) programmatic validation mechanisms to mitigate hallucinations (e.g., syntactic checks and consistency heuristics); and (iv) reproducible metrics that quantify detection rates, extraction accuracy, false positives/negatives, and model behavior across different obfuscation levels. To ensure scalability and allow for reproducibility, the entire workflow is orchestrated through automation tools that record inputs, prompts, raw responses, and validation results.

In the following subsections, we first describe the design and implementation of the Dataset Construction (Section~\ref{sec:obfuscation_process}), including the applied transformations and instrumentation choices. We then detail the AI-assisted Deobfuscation Process (Section~\ref{sec:query_process}), focusing on prompt engineering, output normalization, and the automated evaluation metrics and procedures.

\subsection{Dataset Construction}
\label{sec:obfuscation_process}

The purpose of this process is to build a reproducible and progressively challenging benchmark of programs in which an IoC — in our experiments, an IP address — is systematically concealed using a controlled sequence of obfuscation and cryptographic transformations. 

This stage also focuses on ensuring dataset quality and experimental consistency. Specifically, each transformation is selected to emulate concealment techniques commonly observed in real-world malicious or evasive scripts, while maintaining traceability between the original and transformed samples through structured metadata. 



\paragraph{Transformation pipeline.} 
For each input program, we apply a sequence of transformations arranged into \textbf{12 progressive concealment levels} described in Table~\ref{tab:obfuscation_phases}. These include syntactic manipulation (e.g., variable renaming, string concatenation), structural obfuscation (e.g., control-flow flattening, dead code insertion), and cryptographic encoding (e.g., Base64, XOR, AES-256). This progressive design enables fine-grained evaluation of model robustness across varying obfuscation levels, while ensuring reproducibility through explicit versioning and metadata linkage. Each level modifies the program while preserving its original functional behavior (or, in cases where the transformation intentionally targets only data values, preserving the ability to recover the original IoC given adequate analysis). These levels are intentionally cumulative: higher-numbered levels embed earlier transformations, producing increasingly noisy and semantically layered samples.

\begin{table*}[h]
\caption{Description of the 12 transformation phases for code obfuscation.}
\label{tab:obfuscation_phases}
\centering
\renewcommand{\arraystretch}{1.15}
\setlength{\tabcolsep}{6pt}
\begin{tabular}{p{4.5cm}p{11.5cm}}
\toprule
Phase & Description \\
\midrule
1. Base64 encoding & Replace the plain IoC literal with its Base64 encoding and add the minimal decoding code where needed (thus allowing for straightforward decoding) \\ \hline
2. Identifier obfuscation & Rename variables and functions with non-descriptive identifiers (randomized tokens) and remove comments; do not otherwise alter program logic \\ \hline
3. Dead code injection & Prepend or intersperse non-executed statements and useless functions to increase noise and confuse simplistic pattern matching. \\ \hline
4. Structural obfuscation & Apply an automated process that performs structural transformations, including identifier mangling, control-flow flattening, wrapper functions, and array/table string encoding. \\ \hline
5. XOR encryption & Replace the IoC with an XOR-encrypted byte array and include the small decryption routine and the key within the program (i.e., the program contains a syntactically obvious decrypt function and the key constant). \\ \hline
6. AES-256 encryption & Replace the IoC with an AES-256-encrypted payload and include a decryption routine plus the AES key bytes in the program. This level is intended to represent stronger cryptographic concealment while keeping the sample recoverable without external assistance, since the key is present. \\ \hline
7. XOR + simple obfuscation & Combine the XOR-encrypted IoC (level 6) with identifier renaming and comment removal (level 3). \\ \hline
8. AES + simple obfuscation & Combine AES encryption (level 7) with identifier renaming and comment removal. \\ \hline
9. XOR + dead code & Combine XOR encryption with injected dead code (level 4). \\ \hline
10. AES + dead code &  Combine AES encryption with injected dead code. \\ \hline
11. XOR + advanced obfuscation & Combine XOR encryption with the advanced structural obfuscations of level 5. \\ \hline
12. AES + advanced obfuscation & Combine AES encryption with the advanced structural obfuscations of level 5. \\
\bottomrule
\end{tabular}
\end{table*}

\paragraph{Preservation and verification.} Every new source code obtained through the application of each transformation level undergoes the following automated checks.

\begin{enumerate}
  \item Syntactic sanity. Parse the transformed file with a standard JavaScript parser to ensure the code is syntactically valid.
  \item Behavioral sanity (where applicable). For samples where the original program exposes an execution harness, we execute basic tests to confirm that primary program behavior is preserved (the test harness is minimal to avoid running untrusted code with side effects). When a decryption routine is added, we verify that the provided decryptor returns the original IoC in a controlled execution environment.
  \item Ground-truth validation. This check confirms that the traceability metadata matches the actual embedded representation (for example, decoding Base64 yields the expected IP, running the embedded decryptor returns the stored IoC, etc.).
\end{enumerate}

\paragraph{Rationale for embedding decryption keys.} In obfuscation levels that rely on encryption (i.e., above level 6), the process deliberately includes the decryption routine and key material in the source code. This design choice creates a realistic adversarial scenario often observed in practice (developers or malware authors sometimes keep decryption logic in the same artifact) while ensuring that the ground truth remains recoverable by a capable static/dynamic analysis pipeline. More importantly, it allows to evaluate the model’s ability to (a) recognize cryptographic constructs and decryption patterns, and (b) reason about program-level data flows (rather than measuring brute-force cryptanalysis capability, which is out of scope).

\paragraph{Summary.} The resulting benchmark provides a reproducible, progressively-challenging set of JavaScript artifacts that capture realistic concealment strategies — from trivial modifications, to encrypted IoCs combined with complex structural obfuscation. 

\subsection{AI-assisted deobfuscation}
\label{sec:query_process}

The second component of our methodology is the \textit{AI-assited deobfuscation}, which designs the prompts for the AI models, the rules for normalizing responses, and how the results are evaluated in a reproducible manner. The experiments rely on online, cloud-based LLMs accessed through their corresponding APIs (e.g., OpenAI, Anthropic, Google), as opposed to locally deployed or fine-tuned models. This setup enables consistent querying under controlled parameters while ensuring comparability across systems that differ in architecture and scale. The design goals of this process are three. First, to ensure comparability across models; second, to reduce ambiguities and hallucinations in responses; and third, to automate the evaluation workflow for large-scale experimentation.

\paragraph{Prompting strategy.} Each obfuscated sample is presented to the LLM using a controlled prompt template. We conduct exploratory testing with different LLMs to design the most suitable prompt to achieve our goals. Concretely, the prompt first asks the model to indicate whether an IoC is present or not, restricting to a closed decision with three options (i.e., YES, NO or DON’T KNOW). Second, and if the answer to the previous is YES, it asks for the actual value of the IoC. The prompt wording and examples are kept consistent across all models to minimize prompt-induced variability. To allow for scalable experimentation, queries are issued programmatically through the models' APIs, respecting rate limits, retries, and logging the raw responses for subsequent normalization and analysis. 

The exact prompt used is the following (replacing IOC by the correct type being sought):

\fcolorbox{black}{lightgray}{%
\begin{minipage}{0.9\columnwidth}
\texttt{I am a security analyst and I want to check if this JS code has an [IoC]. I want to know only YES, NO, DON'T KNOW and in case of YES, give me the [IoC]. The results should be ONLY in a JSON with this format example:}%
\end{minipage}%
}

\paragraph{Response normalization.} Since different models vary in verbosity and formatting, raw outputs are normalized into a unified schema:
\begin{itemize}
    \item \textbf{YES / NO / DON’T KNOW}: decision indicating whether an IoC was detected.
    \item \textbf{Extracted value} (only when YES is provided before).
\end{itemize}
Finally, we rely on a post-processing phase to discard irrelevant text, validate whether the provided IoC is valid (using regex patterns), and standardize the output format to allow for the automatic processing of the results.

\paragraph{Summary.}
In summary, the AI-assisted deobfuscation component defines a controlled and reproducible framework for interacting with LLMs. By enforcing a standardized prompting strategy, normalized output schema, and automated API-based execution, the process ensures that responses from different models can be objectively compared. Rather than introducing an entirely new paradigm, this component instantiates and tailors existing LLM evaluation practices to encryption-aware obfuscation and IoC-centric tasks, providing a solid foundation for consistent experimentation across heterogeneous LLM architectures.

\subsection{Validation and Evaluation} In this stage, the normalized responses from the LLMs are automatically validated against the ground-truth metadata collected during the obfuscation process. The evaluation metrics are: 
\begin{itemize}
    \item Detection rate, i.e., the proportion of samples in which the model correctly identifies the presence of an IoC.
    \item Extraction accuracy, i.e., the proportion of samples in which the recovered IP matches the ground truth exactly. It also checks for hallucinations in the answers. 
    \item False positives / false negatives, i.e., the frequency of incorrect YES/NO classifications.
    \item Uncertainty rate, i.e., frequency of DON’T KNOW responses or invalid outputs.
\end{itemize}

\paragraph{Automation pipeline.} The query workflow is fully automated. For each model and each sample, the system:
\begin{enumerate}
    \item Constructs the appropriate prompt template.
    \item Submits the query via API.
    \item Records the raw response and metadata (model, version, temperature, timestamp).
    \item Normalizes and validates the response.
    \item Stores both the raw and normalized results for analysis.
\end{enumerate}

\paragraph{Summary.}
The validation and evaluation process establishes a unified methodology to assess the detection capabilities and reliability of LLMs when analyzing obfuscated code. Through quantitative metrics such as detection accuracy, false positives and negatives, and hallucination rates, this phase measures both the precision and consistency of model reasoning. Automated scripts aggregate results across models and obfuscation levels, enabling large-scale statistical analysis and reproducible benchmarking of LLM performance in adversarial conditions. This pipeline enables large-scale, reproducible experimentation across multiple LLMs.

\section{Experimentation}
\label{sec:experimentation}

In this section, we describe the experimental setup used to evaluate the capability of LLMs to analyze obfuscated code and extract hidden IoCs under different obfuscation levels. We detail the dataset used in our experiments, the model configurations, and the results obtained in terms of detection accuracy, consistency, and robustness against obfuscation and encryption.
To facilitate reproducibility, all experiments are implemented using our open-source Spring Boot framework and scripts, which are publicly available in our GitHub repository~\cite{github-repo}.

\subsection{Dataset}
\label{subsec:experimentation-dataset}

The dataset used in this study was obtained from the open-source JavaScript repository maintained by \textit{The Algorithms Project}~\cite{thealgorithms2024}. This repository contains a large and diverse collection of algorithmic implementations across multiple domains of computer science, including sorting, searching, graph traversal, dynamic programming, string manipulation, mathematics, trees, backtracking, combinatorics, and heuristics. This diversity provides broad coverage in terms of code length, structure, and semantics—an essential factor for evaluating the generalization capacity of Large Language Models (LLMs) when analyzing obfuscated code.

\paragraph{Selection and Preprocessing.}
From the original repository, we curated a subset of 336 JavaScript files. Files were selected to ensure syntactic correctness, absence of external dependencies, and diversity in functional logic. Redundant examples, incomplete implementations, and scripts that could not be processed by the obfuscation pipeline were excluded. Each selected file serves as a baseline for generating progressively obfuscated and encrypted variants within the experimental framework.

\paragraph{Implementation details and instrumentation.}  The obfuscation pipeline is implemented as an automated tool that transforms files in-place and outputs transformed variants with metadata. Key implementation choices include:

\begin{itemize}
  \item Deterministic randomness. All random choices (identifier names, dead-code placements, array permutations, encryption IVs where applicable) use custom seeds, so transformations are reproducible across runs.
  \item Implementation of encryption. XOR encryption is implemented as a small routine that XORs each byte with a key array; for AES-256 we use a reference JavaScript AES implementation (e.g., a vetted AES library \cite{cryptojs} to produce ciphertext and to include an inline decrypt function. In both cases, decryption keys and other cryptographic material are embedded in the source code as a constant variable. This choice 
  prevents introducing an intractable brute-force recovery task.
  \item Obfuscator selection and configuration. For advanced structural obfuscation we employ a well-known JS obfuscation tool (\cite{javascriptobfuscator}). For each transformed sample, we record all the parameters (e.g., control-flow flattening intensity, string array usage) as metadata .
  \item Dead-code templates. Dead code snippets and dummy functions are drawn from a curated template pool to avoid introducing unintended side effects; inserted code is validated to be non-executing or to preserve original semantics.
  \item Insertion policy. For placement of the IoC (or its transformed representation) we use a set of canonical locations: top-level variable, inside a function body, inside an object property, or within a small string table. Distribution across locations is recorded to avoid location bias.
  \item Traceability metadata. For every transformed sample we emit a JSON record containing: original filename, transformation level, applied transformation parameters, inserted IoC canonical form, encryption keys (when embedded), and deterministic seed. This metadata provides the ground truth mapping needed for automatic evaluation and verification.
\end{itemize}

\paragraph{Dataset output and storage.} For each original file, the transformation pipeline produces 12 transformed variants (one per level). Each variant is stored with its corresponding metadata record. All raw inputs, transformed outputs, seeds, and logs are retained in a versioned artifact store to guarantee reproducibility and to allow third parties to re-generate the benchmark artifacts.

\paragraph{Dataset Validation.}
To validate the completeness and representativeness of the dataset, we considered three software metrics proposed in~\cite{aswini2017locmetrics}:
\begin{itemize}
    \item LOC (Lines of Code): used to quantify code size and structural variation across samples.
    \item Function Count: measures modularity and aids in identifying code complexity distribution.
    \item Cyclomatic Complexity: captures the logical branching level, providing a proxy for algorithmic diversity.
\end{itemize}


\begin{table}[t]
\caption{Key dataset features for the original and obfuscated corpora after preprocessing and phase-based transformation.}
\label{tab:dataset_stats}
\centering
\small 
\setlength{\tabcolsep}{4pt} 
\begin{tabular}{lrr}
\toprule
Feature & Original corpus & Obfuscated corpus \\
\midrule
No. files per phase & 336 & 336 \\
Total files (all phases) & 336 & 4,368 \\
No. program categories & 10 & 10 \\
Avg. LOC per file & 112 & 125 \\
LOC range [min, max] & [15, 420] & [15, 433] \\
Avg. functions per file & 1.8 & 2.8 \\
Avg. cyclomatic complexity & 9.3 & 9.5 \\ 
\bottomrule
\end{tabular}
\end{table}

The original corpus exhibits LOC values ranging from 15 to 420 lines per file, with an average of 112 LOC. Function counts vary between 1 and 12 per file, and average cyclomatic complexity remains below 10, providing balanced coverage across different complexity levels. The obfuscated corpus, obtained by applying the 12 transformation phases to the selected files, slightly shifts these distributions, confirming 
that the transformation pipeline preserves a realistic distribution of code sizes and structural properties across phases.

\subsection{Large Language Models (LLMs) Integrated}
The framework allows for querying multiple LLMs programmatically via API through a unified client. Each LLM requires authentication and is accessed via an HTTP-based interface, but all calls are wrapped behind a common abstraction layer that handles request construction, retries, and basic error handling. In our experimentation, we use the models listed in Table~\ref{tab:llm_models}, which cover different providers and architectural families (e.g., GPT, Gemini, Claude).

For each model, the framework normalizes core parameters such as temperature, maximum output length, and system instructions, and enforces a common output schema so that responses can be logged and evaluated in a consistent way. This design enables us to swap or add models with minimal changes to the experimental code, and to run the same deobfuscation and IoC extraction prompts across heterogeneous LLM backends.

A practical limitation of our study is that the evaluated models correspond to commercial releases available at the time of experimentation, and some of them have since been superseded by newer versions. While this may underestimate the absolute performance of the current frontier models, it does not affect the validity of our comparative analysis or the qualitative patterns we observe regarding obfuscation and encryption.

\begin{table}[h]
\caption{Summary of LLMs evaluated in this study.}
\label{tab:llm_models}
\centering
\small
\renewcommand{\arraystretch}{1.2}
\setlength{\tabcolsep}{5pt}
\begin{tabular}{llcr}
\toprule
Model & Provider & Release (Year) & Context Window \\
\midrule
GPT-4 Turbo & OpenAI & 2024 & 128k \\
Gemini 1.5 Pro & Google & 2025 & 1000000 \\
Claude 3 Sonnet & Anthropic & 2024 & 200k \\
Grok-2 & xAI & 2024 & 128k \\
Command-R7B & Cohere & 2024 & 128k \\
\bottomrule
\end{tabular}
\end{table}


\subsection{Results}

This section presents the results of our evaluation. First, we report aggregate detection statistics per model. We then analyze phase-level outcomes across the different obfuscation and encryption stages (P0--P12), showing that increasing concealment intensity hardens the identification of the IoCs.

\begin{table}[t]
    \caption{Overall results per model. DR reflects the detection rate, i.e., ratio of YES an answers. The Acc. reflects the percentage of correct detections among those positive answers. Finally, the \#DK reflects the number of ``Don't Know'' responses.}
    \label{tab:detection_results}
    \centering
    \renewcommand{\arraystretch}{1.2} 
    \begin{tabular}{lrrrr}
        \toprule
        Model & \#Queries & DR & Acc. & \#DK \\
        \midrule
        Anthropic & 4,358 & 38.5\% & 99.7\% & 94 \\
        ChatGPT   & 4,362 & 38.6\% & 99.4\% & 37 \\
        Gemini    & 4,358 & 38.5\% & 87.7\% & 0  \\
        Grok      & 4,357 & 35\% & 100.0\% & 0  \\
        Cohere    & 4,323 & 22.8\% & 100.0\% & 625 \\
        \bottomrule
    \end{tabular}
\end{table}

Table~\ref{tab:detection_results} summarizes the overall detection performance across all phases and samples. We first observe the detection rate. On the one hand, Anthropic, ChatGPT, Gemini, and Grok report a 'YES' in about one third of the queries, with little `Don't Known' responses. On the other hand,  Cohere is more conservative, returning ``Don't Know'' in 625 cases and a DR of 22.8\%.   When they detect an IoC, Anthropic, ChatGPT, Grok, and Cohere maintain near-perfect accuracy (above 99\%), while Gemini exhibits a noticeably lower rate (87.7\%) despite issuing a similar number of findings. 

The aggregate view from Table~\ref{tab:detection_results} highlights heterogeneous trade-offs between assertiveness and uncertainty, but it does not reveal phase-specific behavior for the different obfuscation phases. Therefore, 
Figure~\ref{fig:phase_chart} provides a visual overview of detection outcomes across all transformation phases (P0 to P12). The pie charts shows the three outcomes: successful detection (green, ``YES''), failed detection (red, ``NO''), and explicit uncertainty (yellow, ``Don't Know''). We next analyze the behavior by groups of phases, separating baseline and syntactic obfuscation (P0--P4), cryptographic obfuscation (P5--P6), and combined obfuscation techniques (P7--P12).

\newcommand{\piechart}[7]{%
  \begin{tikzpicture}[baseline={(0,-0.2)}]
    \def\r{#1}%
    \pgfmathsetmacro{\angA}{#2*3.6}
    \pgfmathsetmacro{\angB}{#3*3.6}
    \pgfmathsetmacro{\angC}{#4*3.6}
    \pgfmathsetmacro{\angAB}{\angA+\angB}
    \fill[#5] (0,0) -- (0:\r) arc (0:\angA:\r) -- cycle;
    \fill[#6] (0,0) -- (\angA:\r) arc (\angA:\angAB:\r) -- cycle;
    \fill[#7] (0,0) -- (\angAB:\r) arc (\angAB:360:\r) -- cycle;
    \draw[line width=0.4pt] (0,0) circle (\r);
  \end{tikzpicture}%
}

\begin{figure}[ht]
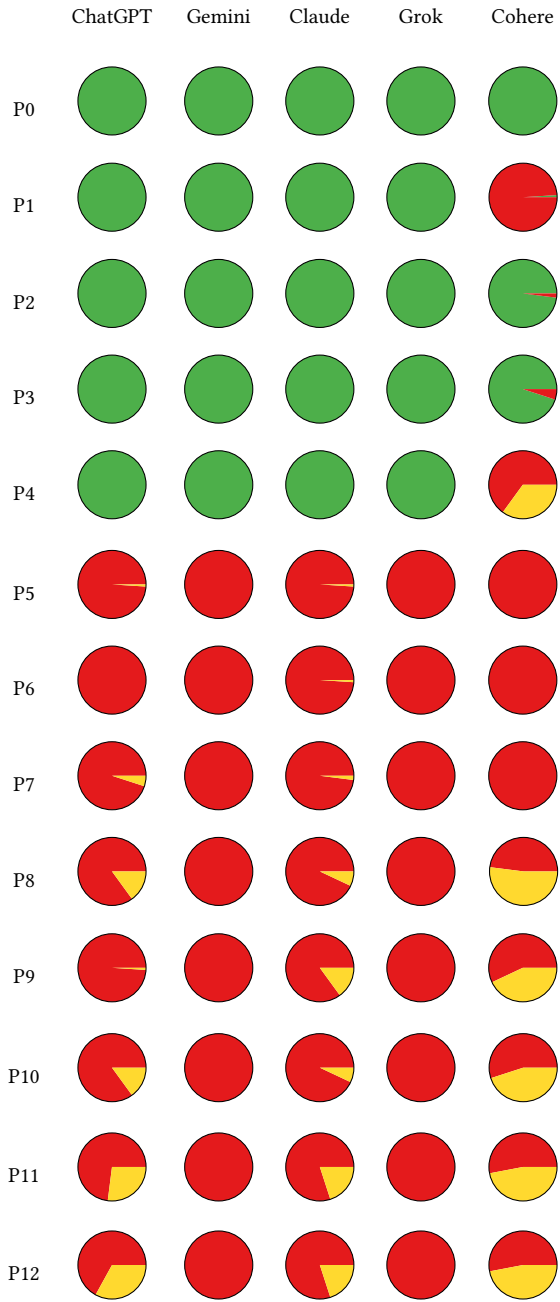

\centering
\small
\setlength{\tabcolsep}{6pt}

\begin{tabular}{cccccc}
   & ChatGPT & Gemini & Claude & Grok & Cohere \\\\[1ex]
P0 &
\piechart{0.45cm}{100}{0}{0}{yesgreen}{nored}{dkyellow} &
\piechart{0.45cm}{100}{0}{0}{yesgreen}{nored}{dkyellow} &
\piechart{0.45cm}{100}{0}{0}{yesgreen}{nored}{dkyellow} &
\piechart{0.45cm}{100}{0}{0}{yesgreen}{nored}{dkyellow} &
\piechart{0.45cm}{100}{0}{0}{yesgreen}{nored}{dkyellow} \\\\

P1 &
\piechart{0.45cm}{100}{0}{0}{yesgreen}{nored}{dkyellow} &
\piechart{0.45cm}{100}{0}{0}{yesgreen}{nored}{dkyellow} &
\piechart{0.45cm}{100}{0}{0}{yesgreen}{nored}{dkyellow} &
\piechart{0.45cm}{100}{0}{0}{yesgreen}{nored}{dkyellow} &
\piechart{0.45cm}{1}{99}{0}{yesgreen}{nored}{dkyellow} \\\\

P2 &
\piechart{0.45cm}{100}{0}{0}{yesgreen}{nored}{dkyellow} &
\piechart{0.45cm}{100}{0}{0}{yesgreen}{nored}{dkyellow} &
\piechart{0.45cm}{100}{0}{0}{yesgreen}{nored}{dkyellow} &
\piechart{0.45cm}{100}{0}{0}{yesgreen}{nored}{dkyellow} &
\piechart{0.45cm}{98}{2}{0}{yesgreen}{nored}{dkyellow} \\\\

P3 &
\piechart{0.45cm}{100}{0}{0}{yesgreen}{nored}{dkyellow} &
\piechart{0.45cm}{100}{0}{0}{yesgreen}{nored}{dkyellow} &
\piechart{0.45cm}{100}{0}{0}{yesgreen}{nored}{dkyellow} &
\piechart{0.45cm}{100}{0}{0}{yesgreen}{nored}{dkyellow} &
\piechart{0.45cm}{95}{5}{0}{yesgreen}{nored}{dkyellow} \\\\

P4 &
\piechart{0.45cm}{100}{0}{0}{yesgreen}{nored}{dkyellow} &
\piechart{0.45cm}{100}{0}{0}{yesgreen}{nored}{dkyellow} &
\piechart{0.45cm}{100}{0}{0}{yesgreen}{nored}{dkyellow} &
\piechart{0.45cm}{100}{0}{0}{yesgreen}{nored}{dkyellow} &
\piechart{0.45cm}{0}{65}{35}{yesgreen}{nored}{dkyellow} \\\\

P5 &
\piechart{0.45cm}{0}{99}{1}{yesgreen}{nored}{dkyellow} &
\piechart{0.45cm}{0}{100}{0}{yesgreen}{nored}{dkyellow} &
\piechart{0.45cm}{0}{99}{1}{yesgreen}{nored}{dkyellow} &
\piechart{0.45cm}{0}{100}{0}{yesgreen}{nored}{dkyellow} &
\piechart{0.45cm}{0}{100}{0}{yesgreen}{nored}{dkyellow} \\\\

P6 &
\piechart{0.45cm}{0}{100}{0}{yesgreen}{nored}{dkyellow} &
\piechart{0.45cm}{0}{100}{0}{yesgreen}{nored}{dkyellow} &
\piechart{0.45cm}{0}{99}{1}{yesgreen}{nored}{dkyellow} &
\piechart{0.45cm}{0}{100}{0}{yesgreen}{nored}{dkyellow} &
\piechart{0.45cm}{0}{100}{0}{yesgreen}{nored}{dkyellow} \\\\

P7 &
\piechart{0.45cm}{0}{95}{5}{yesgreen}{nored}{dkyellow} &
\piechart{0.45cm}{0}{100}{0}{yesgreen}{nored}{dkyellow} &
\piechart{0.45cm}{0}{98}{2}{yesgreen}{nored}{dkyellow} &
\piechart{0.45cm}{0}{100}{0}{yesgreen}{nored}{dkyellow} &
\piechart{0.45cm}{0}{100}{0}{yesgreen}{nored}{dkyellow} \\\\

P8 &
\piechart{0.45cm}{0}{85}{15}{yesgreen}{nored}{dkyellow} &
\piechart{0.45cm}{0}{100}{0}{yesgreen}{nored}{dkyellow} &
\piechart{0.45cm}{0}{93}{7}{yesgreen}{nored}{dkyellow} &
\piechart{0.45cm}{0}{100}{0}{yesgreen}{nored}{dkyellow} &
\piechart{0.45cm}{0}{48}{52}{yesgreen}{nored}{dkyellow} \\\\

P9 &
\piechart{0.45cm}{0}{99}{1}{yesgreen}{nored}{dkyellow} &
\piechart{0.45cm}{0}{100}{0}{yesgreen}{nored}{dkyellow} &
\piechart{0.45cm}{0}{85}{15}{yesgreen}{nored}{dkyellow} &
\piechart{0.45cm}{0}{100}{0}{yesgreen}{nored}{dkyellow} &
\piechart{0.45cm}{0}{57}{43}{yesgreen}{nored}{dkyellow} \\\\

P10 &
\piechart{0.45cm}{0}{85}{15}{yesgreen}{nored}{dkyellow} &
\piechart{0.45cm}{0}{100}{0}{yesgreen}{nored}{dkyellow} &
\piechart{0.45cm}{0}{93}{7}{yesgreen}{nored}{dkyellow} &
\piechart{0.45cm}{0}{100}{0}{yesgreen}{nored}{dkyellow} &
\piechart{0.45cm}{0}{55}{45}{yesgreen}{nored}{dkyellow} \\\\

P11 &
\piechart{0.45cm}{0}{73}{27}{yesgreen}{nored}{dkyellow} &
\piechart{0.45cm}{0}{100}{0}{yesgreen}{nored}{dkyellow} &
\piechart{0.45cm}{0}{80}{20}{yesgreen}{nored}{dkyellow} &
\piechart{0.45cm}{0}{100}{0}{yesgreen}{nored}{dkyellow} &
\piechart{0.45cm}{0}{53}{47}{yesgreen}{nored}{dkyellow} \\\\

P12 &
\piechart{0.45cm}{0}{67}{33}{yesgreen}{nored}{dkyellow} &
\piechart{0.45cm}{0}{100}{0}{yesgreen}{nored}{dkyellow} &
\piechart{0.45cm}{0}{80}{20}{yesgreen}{nored}{dkyellow} &
\piechart{0.45cm}{0}{100}{0}{yesgreen}{nored}{dkyellow} &
\piechart{0.45cm}{0}{53}{47}{yesgreen}{nored}{dkyellow} \\\\
\end{tabular}

\caption{LLM detection outcomes across all obfuscation phases (P0--P12). Green sections indicate successful detections, red sections indicate missed detections, and yellow sections indicate explicit ``Don't Know'' responses.}
\label{fig:phase_chart}
\end{figure}

\subsubsection{Baseline and Syntactic Obfuscation (P0--P4)}

The baseline scenario (P0) serves as a baseline: all evaluated LLMs achieve 100\% detection when the IoC appears in plain text, confirming their capability to identify and extract IoCs in source code. As obfuscation techniques progress through simple transformations—P1 (Base64 encoding), P2 (identifier obfuscation), P3 (dead code injection), and P4 (advanced structural obfuscation)—all most models perform well. Indeed, ChatGPT, Claude, Gemini and Grok achieve 100\% across all four phases without reporting uncertainties. Cohere, however, shows progressive degradation: 1\% missed detections in P1, 2\% in P2, 5\% in P3, and in P4, 65\% misses plus 35\% explicit ``Don't Know'' responses. This suggests that while simple syntactic obfuscation does not impede most modern LLMs, more complex structural transformations begin to degrade Cohere's recognition capabilities. Recall from Table~\ref{tab:detection_results} that in most cases, when the response is `YES', the actual IoC is accurately reported. 

\subsubsection{Cryptographic Obfuscation (P5--P6)}

The use of encryption techniques produces a sharp performance drop across all models. In P5 (XOR encryption with the key embedded in the code), all LLMs transition from systematic success to near-complete failure: the pies in Figure~\ref{fig:phase_chart} show that positive detections virtually disappear, and the outcome mass moves to missed detections (red) and, in some cases, explicit uncertainty (yellow). A similar pattern emerges in P6 (AES-based encryption): for all models, encrypted IoCs become effectively opaque, regardless of whether the underlying primitive is XOR or AES. This transition from 100\% success in P4 to near-zero detection in P5 and P6 indicates that encryption, even when the decryption key is present in the source, fundamentally disrupts the token- and pattern-based mechanisms that LLMs rely on for IoC recognition.

\subsubsection{Combined Obfuscation Techniques (P7--P12)}

Phases P7 through P12 combine encryption techniques with the previous obfuscation layers: simple identifier renaming (P7 and P8), dead code injection (P9 and P10), and advanced structural obfuscation (P11 and P12). Overall, the results are similar to those with pure encryption phases: once the code is encrypted, detection of IoCs remains essentially zero for all models, and the added syntactic or structural layers do not produce a systematic additional degradation. For example, in P8 and P10, the pies remain dominated by red (failures) and yellow (uncertainty) segments, indicating that encryption is the dominant failure mode and that extra obfuscation only induces minor shifts in the balance between “NO” and “Don't Know”.

In the most complex scenarios, P11 (XOR + advanced obfuscation) and P12 (AES + advanced obfuscation), there is still no genuine recovery of detection capability: models simply redistribute part of the failures into explicit ``Don't Know'' responses, but no green success slices appear. This suggests that certain structural combinations may cause some LLMs to become slightly more cautious (increasing uncertainty), yet they still fail to recover IoCs once these have been concealed through encryption.

\paragraph{Response Consistency and Hallucinations.}

Model behavior also varied significantly in terms of response format and reliability. While some models adhered closely to the expected JSON schema, others produced inconsistent or incomplete outputs, affecting automated evaluation. Hallucinations — cases where a model reported an IP address that did not exist in the code — are also observed in some minor cases, reinforcing the need for strict response validation in security-critical applications.

To quantify this phenomenon, we computed the proportion of hallucinations per model as the ratio of false positives to total queries. Table~\ref{tab:hallucination_results} presents hallucination rates across all five models and all phases combined.

\begin{table}[t]
    \caption{Hallucination rate observed across LLMs across all phases. While aggregate rates remain below 5\%, their non-trivial presence in cybersecurity workflows is significant.}
    \label{tab:hallucination_results}    \centering
    \begin{tabular}{lr}
        \toprule
        \textbf{Model} & \textbf{Hallucination Rate (\%)} \\
        \midrule
        Anthropic & 0.11 \\
        ChatGPT & 0.23 \\
        Gemini & 4.8 \\
        Grok & 0 \\
        Cohere & 0 \\
        \bottomrule
    \end{tabular}
\end{table}

Although the aggregate hallucination rates in Table~\ref{tab:hallucination_results} remain low—with Anthropic and ChatGPT under 0.25\%, and Grok and Cohere at 0\%—Gemini's 4.8\% rate is non-negligible in cybersecurity contexts where false indicators can significantly distort automated threat intelligence and trigger unnecessary incident response procedures, potentially affecting benign endpoints. Table~\ref{tab:hallucination_examples} provides a detailed taxonomy of hallucinated IoCs, revealing systematic patterns rather than random noise.

\begin{table}[t]
\caption{Main observed hallucinations and examples.}
\label{tab:hallucination_examples}
\centering
\renewcommand{\arraystretch}{1.15}
\begin{tabular}{llr}
\toprule
Hallucinated type & Examples & Count \\
\midrule
IP addresses & 192.168.17.101 & 206 \\
& 192.168.1.1 & 4 \\
& 198.51.100.1 & 1 \\
& 160.9.34.0 & 1 \\
& 34.207.87.0 & 1 \\
& 8.7.7.8 & 1 \\
& 172.31.0.0/16 & 1 \\
& 192.168.17.105 & 1 \\
& 72.70.83.9 & 1 \\
& 185.199.108.153 & 1 \\
Domain & www.cs.auckland.ac.nz & 1 \\
& en.wikipedia.org & 1 \\
String & Not Found in Plain Text & 1 \\
& N/A & 2 \\
& Encrypted data & 1 \\
Base64 & U2Fsd [...] l5ZY= & 1 \\
\bottomrule
\end{tabular}
\end{table}

A closer inspection of the hallucination distribution reveals striking systematic biases. The private IP address \texttt{192.168.17.101} accounts for 206 out of the recorded hallucinations, representing over 95\% of the total false positives. This overwhelming concentration suggests that models—when unable to decode cryptographic or obfuscated material—revert to a learned heuristic: private IP ranges serve as a plausible ``default'' response when confronted with incomplete or ambiguous security contexts. In some cases, models may partially decode encrypted material (e.g., interpreting certain bytes as numeric components), leading to the generation of private IPs that share structural similarity with the true values but are fundamentally incorrect.

Beyond the dominant private IP bias, additional patterns emerge. Even if we instruct in the prompt to extract IP values, a few domain names appear among the hallucinations (such as auckland.ac.nz and wikipedia.org variants), alongside domain-like encrypted strings, suggesting that models generalize from training data where security contexts frequently mention external resources. Similarly, the sporadic appearance of Base64-encoded strings (e.g., \texttt{U2FsdGVkX1[..]}) indicates overgeneralization: the model recognizes the encoding format but fails to isolate the true IoC, returning the encrypted representation itself as a purported ``finding.''

These patterns indicate that hallucinations are not purely stochastic but emerge from systematic contextual overfitting and probabilistic reasoning typical of generative models. When presented with obfuscated material, LLMs apply learned associations rather than principled inference, leading to predictable and repeatable errors. In practical terms, this means that LLMs may generate credible yet incorrect indicators when confronted with encrypted or heavily obfuscated code, amplifying the operational risk in automated IoC detection pipelines.

\subsection{Analysis of Research Questions}

We finalize the evaluation section by revisiting and discussing the proposed RQs. 

\subsubsection{RQ1: Are LLMs capable of finding security-relevant artifacts in obfuscated source code?}

The results provide a affirmative answer to RQ1: LLMs show robust capability to extract security-relevant artifacts (specifically IP addresses) from source code when IoCs are represented in plain text, and when code is subject to common syntactic obfuscation. Across the baseline and phases P0 through P4, four of five models (ChatGPT, Claude, Gemini, Grok) maintained detection rates at or near 100\%, indicating that their transformer-based architectures have internalized strong pattern recognition for security-relevant literals and semantically transparent transformations.

However, this capability is bounded. Cohere's progressive degradation through P1--P4, culminating in 65\% missed detections and 35\% explicit uncertainty in P4, demonstrates that even among advanced LLMs, architectural differences and training data can significantly influence robustness under syntactic stress. More critically, the performance cliff observed at P5 (the introduction of any cryptographic transformation) reveals a fundamental limitation: LLMs trained on code corpora cannot reliably extract encrypted IoCs even when the decryption key and routine are syntactically present within the code. This suggests that LLMs primarily rely on surface-level token recognition and semantic pattern matching rather than reasoning about program semantics, control flow, or cryptographic operations.

Accordingly, RQ1 should be answered conditionally: \textit{LLMs are capable of finding security artifacts in non-encrypted obfuscated code with high accuracy, but they fail decisively when confronted with cryptographic obfuscation, regardless of whether decryption keys are present in the code.}

\subsubsection{RQ2: How do advanced obfuscation techniques, such as encryption, affect the ability of LLMs to accurately extract security-relevant artifacts?}

RQ2 directly addresses the magnitude and nature of performance degradation under increasing obfuscation complexity. The data reveal three critical insights:

\begin{enumerate}
    \item \textbf{Cryptography as a hard boundary:} The transition from P4 (advanced structural obfuscation with 100\% success for most models) to P5 (XOR encryption with embedded key) results in complete performance degradation. ChatGPT, Claude, and Anthropic drop from 100\% to 99\% failure rates. This abrupt cliff—rather than gradual degradation—indicates that encryption introduces a qualitatively different failure mode than syntactic obfuscation. Models do not progressively ``lose signal'' as transformations deepen; instead, they encounter a fundamental boundary where token-level analysis becomes inadequate.

    \item \textbf{Encryption strength is not the limiting factor:} Both XOR (P5, P6) and AES-256 (P7, P8) produce nearly identical failure rates across all models (>95\% misses for most, excluding Grok's anomalous 100\% success). This parity suggests that encryption strength—the computational difficulty of breaking the cipher without external keys—is orthogonal to the LLM failure mode. Rather, the presence of encrypted material itself (regardless of cipher strength or key inclusion) prevents models from reconstructing the original plaintext. The inclusion of decryption routines and keys in the source code does not enable recovery, suggesting that LLMs lack the capacity to reason about cryptographic operations symbolically or to trace data flow through encryption routines.

    \item \textbf{Additional obfuscation layers do not significantly amplify failure:} Combining encryption with dead code injection (P9, P10) or structural obfuscation (P11, P12) does not substantially worsen performance beyond the baseline encryption-induced failure. This is particularly evident in the minor performance variance in P11 and P12, where Claude shows 80\% misses and ChatGPT 67\%--73\%, compared to the near-100\% failures in P5--P8. This suggests that once encryption is present, additional syntactic noise provides no further degradation—and in some cases, may even permit slight recovery as models apply more intensive reasoning to complex structural transformations.
\end{enumerate}

In synthesis, RQ2 reveals that \textit{cryptographic obfuscation represents a qualitative threshold beyond which LLM extraction capability effectively ceases, whereas simpler obfuscation techniques (identifier renaming, dead code, structural flattening) have minimal impact on models already trained on semantically rich code corpora. The effect of encryption is binary: presence or absence, not a matter of degree.}

\section{Discussion}
\label{sec:discussion}

The empirical results of this study provide a novel understanding of the strengths and limitations of large language models (LLMs) for detecting security-relevant artifacts—specifically IoCs, such as IP addresses—within source code subjected to varied levels of obfuscation.

The findings support that LLMs are highly effective in identifying artifacts when these appear in plaintext or have only undergone simple syntactic modifications. This is evidenced by near-perfect accuracy in the baseline and first obfuscation phases, where models like ChatGPT, Claude, and Gemini demonstrate robust pattern recognition capabilities. Such performance confirms their utility for static code analysis and initial triage tasks when threat indicators are explicit.

However, detection capability decreases drastically in the presence of advanced obfuscation, particularly when cryptographic methods such as XOR or AES-256 encryption are introduced—even when decryption routines and keys are embedded in the code. LLMs do not attempt or succeed in decoding; their approach is limited to direct pattern matching, revealing a fundamental barrier: current LLM architectures lack true semantic reasoning and transformation capabilities. They are unable to symbolically interpret or simulate code execution required for deobfuscation in these scenarios.

The results further highlight the importance of query design and validation, as ambiguous instructions increase the risk of hallucinated outputs: fabricated IPs and false positives that, while low in occurrence, present operational risks in security automation. Also, the results can be affected by the emergence of new AI models that can deal with the aforementioned limitations. In this regard, the scientific framework presented in this paper can serve to conduct future research, including novel obfuscation techniques and new models.


\section{Conclusions}
\label{sec:conclusion}

This work demonstrates that current large language models excel at identifying Indicators of Compromise when these are presented in clear text or with limited syntactic obfuscation. Their accuracy and speed make them promising as support tools in cybersecurity tasks involving code review and automated threat detection. 

Nonetheless, substantial limitations persist. The evaluated LLMs are unable to effectively extract security artifacts when obfuscation escalates to cryptographic concealment, regardless of the presence of decryption methods and keys within the codebase. The inability to symbolically decode or semantically reason across code transformations marks a critical boundary in their applicability. Additionally, while hallucination rates remain statistically low, any instance of a fabricated or incorrect IoC highlights reliability concerns in automated security workflows.

Overall, these findings call for cautious integration of LLMs into cybersecurity pipelines, ideally as part of hybrid frameworks where traditional static or dynamic analysis augments their capabilities. Continued research into improving contextual understanding, reasoning, and uncertainty management within LLM architectures is essential for tackling advanced threat detection use cases.


\nocite{thealgorithms2024,
        aswini2017locmetrics,
        patsakis2024assessing,
        shao2024empirical,
        xu2020layered,
        rauti2024enhancing,
        ieeeTBD}
\bibliographystyle{ACM-Reference-Format}   
\bibliography{references}

\end{document}